\documentclass{amsart}
\usepackage{amssymb}
\usepackage[hyperindex=true]{hyperref}
\mathchardef\ordinarycolon\mathcode`\:     
\mathcode`\:=\string"8000
\def\vcentcolon{\mathrel{\mathop\ordinarycolon}} \begingroup
\catcode`\:=\active \lowercase{\endgroup \let :\vcentcolon }

\newcommand{\Z}{{\mathbb Z}}  
\newcommand{\C}{{\mathbb C}} 
\newcommand{\ket}[1]{|#1\rangle} 
\newcommand{\e}{\mathrm{e}}
\renewcommand{\i}{\mathrm{i}}
\newcommand{\poly}{\mathrm{poly}}

\begin{document}

\title[Quantum Computing Discrete Logarithms\dots]{Quantum Computing Discrete Logarithms \\
with the Help of a Preprocessed State}
\author{Wim van Dam}
\address{Massachusetts Institute of Technology, 
Center for Theoretical Physics,
77 Massachusetts Avenue, 
Cambridge, MA 02139-4307, USA}
\thanks{Report no.~MIT-CTP 3446}  
\email{vandam@mit.edu}


\begin{abstract}
An alternative quantum algorithm for the discrete logarithm problem is
presented.  The algorithm uses two quantum registers and two Fourier
transforms whereas Shor's algorithm requires three registers and four
Fourier transforms.  A crucial ingredient of the algorithm is a
quantum state that needs to be constructed before we can perform the
computation.  After one copy of this state is created, 
the algorithm can be executed arbitrarily many times.
\end{abstract}

\maketitle

\section{Introduction}
In 1994, Peter Shor described an efficient, polynomial time, 
quantum algorithm for the discrete logarithm problem\cite{shor}.
Shor's protocol is based on the period finding capability
of quantum computers and its initial version was a 
probabilistic algorithm.  Following this work,
several authors have presented exact versions of
Shor's algorithm \cite{BH,MZ}, 
based on the method of `amplitude amplification'. 

Here we present an alternative algorithm for the discrete 
logarithm. The algorithm requires the preprocessing of
a state that is specific for the group $G$ and its 
generator $g$ for which we want to calculate the 
discrete logarithm.  The size of this `chi state'
is $\log |G|$ qubits and it can be created efficiently
with zero error probability.
The actual discrete logarithm algorithm is 
more efficient than Shor's version and because the
chi state can be reused indefinitely, we can view 
the production of it as a form of `preprocessing' 
that is especially worthwhile if we intend to solve 
many instances of the discrete logarithm problem for 
a fixed group.
Provided that we have a perfect version of the chi 
state and we can perform the quantum Fourier transform 
over $\Z/m\Z$ (with $m$ the order of the group $G$)
exactly, the algorithm presented here is deterministic.
Furthermore it is possible to perfectly copy the $\chi$
state, hence after one quantum computer has produced 
the state, other computers can acquire the state
with at no extra cost.  Typically, $G$ would be 
the multiplicative `mod $n$' group $(\Z/n\Z)^\times$
with $\phi(n)=m$, but the algorithm works for every
cyclic group $G$.

The reader is referred to \cite{NC} for an introduction
in the theory of quantum computation. 
Throughout the text we assume that we can perform the 
quantum Fourier transform over the additive group $\Z/m\Z$
exactly; see \cite{MZ} for when and how this can be done.

\section{The Algorithm}
Let $G$ be the multiplicative group of order $m$ generated
by $g$ such that $G=\{g^1,g^2,\dots,g^m=1\}$.  For a fixed $g$, 
the \emph{discrete logarithm problem} is to determine the 
power $p \in \Z/m\Z$ of a given element $g^p\in G$
(we use the notation $\log_g(g^p) := p$).  Throughout the
article we assume that the order $m$ is known.

As mentioned in the introduction, the algorithm consists
of two parts: the preprocessing of a `chi state' and the 
actual algorithm, which can be executed arbitrarily many times
on one copy of the chi state.  Before we describe these 
two parts of the algorithm, we will define some of its 
ingredients, which are also used in Shor's algorithm.
\begin{description}
\item[Fourier transform]
For the additive group $\Z/m\Z$ the quantum Fourier transform $F$,
which is a unitary operation, is defined by 
\begin{eqnarray*}
F:\ket{x} & \longmapsto & \frac{1}{\sqrt{m}}\sum_{y=0}^{m-1}{\zeta_m^{xy}\ket{y}},
\end{eqnarray*}
for all $x\in\Z/m\Z$ and $\zeta_m := \e^{2\pi \i/m}$.
How to efficiently implement the Fourier transform in circuits of
size $\poly(\log m)$ is explained in, for example, \cite{revisited}.
 For which $m$ we can implement $F$ exactly and 
how is discussed in, for example, \cite{MZ}.
\item[Division operator] We assume that multiplication and division 
in $G$ can be done efficiently (in time $\poly(\log m)$), and hence 
using repeated powering $x\mapsto x^2\mapsto x^4\cdots$, we
can efficiently calculate any power $x^r$ for $-m \leq r \leq m$.
This shows that the following two reversible `division operators' 
can be implemented efficiently as well:
\begin{eqnarray*}
D^\alpha:\ket{x,y} & \longmapsto & \ket{x,y/x^\alpha}, \\
D_x:\ket{\alpha,y} & \longmapsto & \ket{\alpha,y/x^\alpha},
\end{eqnarray*}
for all $x,y\in G$ and $\alpha\in \Z/m\Z$.
\end{description}

We are now ready to describe the two parts of the quantum algorithm.
First, in \S\ref{sec:chi}, we will define the `chi state', which 
is crucial for the algorithm. We will mention some of its properties
and show the state can be prepared in an efficient way.
After that, in \S\ref{sec:alg}, the actual algorithm will be given.

\subsection{The Chi State, Its Properties and Its Preparation}\label{sec:chi}
Given $g$ and the group $G$, define the chi state by
\begin{eqnarray*}
\ket{\chi} & := & \frac{1}{\sqrt{m}}\sum_{r=0}^{m-1}{\zeta_m^{r}\ket{g^r}}.
\end{eqnarray*} 
We use the symbol $\chi$ for this state because its phase
values $\zeta^r_m$ are the values of the multiplicative 
character $\chi:G\rightarrow \C$ with $\chi(g^r) := \zeta_m^r$
for all $r\in\Z/m\Z$ and hence with $\chi(xy) = \chi(x)\chi(y)$.  
For every $\alpha \in \Z/m\Z$ we also
define the $\alpha$-th power of the chi state by 
\begin{eqnarray*}
\ket{\chi^\alpha} & := & \frac{1}{\sqrt{m}}\sum_{r=0}^{m-1}{\zeta_m^{\alpha r}\ket{g^r}}.
\end{eqnarray*}
Note that $\ket{\chi^0}$ is the uniform superposition
of the elements of $G$.

Using the $D^\alpha$ operation chi states can be copied to 
arbitrary $\chi^\alpha$ states.
It is straightforward to check that if we apply a $D^\alpha$ operation 
to a state $\ket{g^s}\ket{\chi}$ we will induce the phase change 
$\ket{g^s,\chi}\mapsto\zeta_m^{\alpha s}\ket{g^s,\chi}$.
Hence, if we apply $D^\alpha$ to a uniform superposition 
of $G$ and a $\chi$-state, we obtain a new $\chi^\alpha$
state without losing the original $\ket{\chi}$:
\begin{eqnarray*}
D^\alpha: \frac{1}{\sqrt{m}}\sum_{x\in G}{\ket{x}\ket{\chi}}
& \longmapsto & \ket{\chi^\alpha}\ket{\chi}.
\end{eqnarray*}
In general we have in fact the mapping 
$D^\alpha:\ket{\chi^\beta}\ket{\chi^\gamma}
\mapsto \ket{\chi^{\beta+\alpha\gamma}}\ket{\chi^\gamma}$.
Under the assumption that it is easy to create the 
uniform superposition $\ket{\chi^0}$, we thus see that 
we can efficiently create arbitrary $\ket{\chi^\alpha}$ states,
as soon as we have an initial state $\ket{\chi}$.
To create the first chi state, we use the following
zero error procedure.

\subsection*{Chi State Preparation Algorithm:}
\textsl{Let $g$ be the generator
of the group $G = \{g,g^2,\dots,g^m=g^0=1\}$.
\begin{enumerate}
\item Initialize two $\log m$ qubit registers to $\ket{0,0}$ and apply the 
Fourier transform over $\Z/m\Z$ to the left one. Next, 
calculate in the right register powers $g^r$ where the exponent
$r$ is read from the left register.  This step gives the 
transformation
\begin{eqnarray*}
\ket{0,0} & \longmapsto & \frac{1}{\sqrt{m}}\sum_{r=0}^{m-1}{\ket{r,g^r}}.
\end{eqnarray*}
 \item Apply the Fourier transform over $\Z/m\Z$ to the first register:
\begin{eqnarray*}
F\otimes I: \frac{1}{\sqrt{m}}\sum_{r=0}^{m-1}{\ket{r,g^r}}
& \longmapsto & 
\frac{1}{m}\sum_{s,r=0}^{m-1}{\zeta_m^{rs}\ket{s,g^r}}.
\end{eqnarray*}
Note that this state equals $\sum_s {\ket{s,\chi^s}}/\sqrt{m}$.
\item Measure the $s$-register.  If $\gcd(s,m)\neq 1$, go back to 
step $1$ and repeat the protocol.  
Otherwise, continue with the state $\ket{s,\chi^s}$.
\item Clear the $s$ register and replace it with the 
uniform superposition of elements of $G$ such that we
obtain the state $\ket{\chi^0,\chi^s}$. 
\item Apply $D^{(1/s)}$ (as $1/s := s^{-1}$ is well-defined in $\Z/m\Z$),
such that we get the transformation 
$\ket{\chi^0,\chi^s} \mapsto \ket{\chi^1,\chi^s}$. 
Remove the right register, yielding $\ket{\chi}$.
\end{enumerate}}

All steps in the above algorithm can be done in time $\poly(\log m)$.  
The probability that the observed $s$ in Step~1 is co-prime with $m$
is $\phi(m)/m$, which is lower bounded by $\Omega(1/\log(\log m))$. 
Hence the expected number of times that we have to repeat the 
algorithm until we reach Step~4 is $O(\log(\log m))$.
In all, and assuming that we can perform the Fourier transform 
exactly, this shows that this algorithm produces the state 
$\ket{\chi}$ with zero error probability and has expected running
time $\poly(\log m)$.  
Using amplitude amplification \cite{BH} and
knowledge about $\phi(m)$ we could make this algorithm exact,
but because we need to prepare $\ket{\chi}$ only once, we 
do not bother.  (Note again that copying the $\chi$-state via the 
operation $D^1:\ket{\chi^0,\chi} \mapsto \ket{\chi,\chi}$ is 
deterministic and more simple than the just described chi preparation
algorithm.)

\subsection{Using the Chi State for the Discrete Logarithm Problem}\label{sec:alg}
The crucial property of the chi state that we will use
in the logarithm algorithm is its phase changing behavior
when we apply $D_x$ to it.
Given an element $x=g^p \in G$, the $D_x$ transform on 
$\ket{\alpha}$ and $\ket{\chi}$ has the following effect 
(which is shown with the help of the equality $\sum_r {\zeta^r_m\ket{g^r/g^{p\alpha}}} = 
\sum_r{\zeta_m^{\alpha p+r}\ket{g^r}}$):
\begin{eqnarray*}
D_x:\ket{\alpha}\ket{\chi} &\longmapsto& \zeta_m^{\alpha p}\ket{\alpha}\ket{\chi},
\end{eqnarray*}
with $p := \log_g(x)$.
This `multiplicative phase kick-back trick' (cf.~\cite{revisited} for
the additive version) is used to 
calculate  $\log_g x$ in the following algorithm.

\subsection*{Discrete Logarithm Algorithm:}
\textsl{Given the generator $g$, a state $\ket{\chi}$ and the 
input value $x=g^p$, perform the following $3$ steps.
\begin{enumerate}
\item{Create a uniform superposition of $\alpha$'s by 
applying the Fourier transform over $\Z/m\Z$ to $0$:
\begin{eqnarray*}
\ket{0} & \longmapsto & \frac{1}{\sqrt{m}}\sum_{\alpha=0}^{m-1}{\ket{\alpha}}.
\end{eqnarray*}}
\item{With the $\chi$ state as the second register, apply the $D_x$ transform 
to this superposition, giving:
\begin{eqnarray*}
D_x:\frac{1}{\sqrt{m}}\sum_{\alpha=0}^{m-1}{\ket{\alpha}}\ket{\chi}
& \longmapsto & 
\frac{1}{\sqrt{m}}\sum_{\alpha=0}^{m-1}{\zeta_m^{\alpha \log_g(x)}\ket{\alpha}}\ket{\chi}.
\end{eqnarray*}
}
\item{Recover the logarithm $p$ by applying an inverse Fourier transform 
(over $\Z/m\Z$) to the first register, yielding the final state 
$\ket{\log_g(x)}\ket{\chi}$.}
\end{enumerate}}

The complexity of the algorithm consists of two Fourier transforms 
over $\Z/m\Z$ and one implementation of $D_x$, which can all
be done in time $\poly(\log m)$.
 If these transformations
are performed  perfectly and the state $\chi$ is exact, 
then the above algorithm finds the discrete logarithm $\log_g(x)$
with probability $1$.  Note also that the chi state did not get
destroyed in the computation, and hence can be reused.
 
\section{Discussion}
The two $\log m$ qubit registers and two Fourier transforms
over $\Z/m\Z$ of the above algorithm are improvements over
the exact version of Shor's algorithm, as described in \cite{MZ}, 
which requires three quantum registers of $\log m$ qubits and 
four Fourier transforms.  Also the `exactness' of this algorithm
is more straightforward as we did not need to use amplitude 
amplification \cite{BH} to suppress the errors.

If we allow measurements with classical interactions during 
the computation, we can use the semi-classical Fourier transform
over $\Z/2^k\Z$ \cite{GN} 
to reduce the size of the first register to one coherent qubit. 
By taking $k\approx \log m$,  the above algorithm gives a probabilistic 
procedure with $\log m$ measurements during its Fourier 
transform, while the standard semi-classical discrete logarithm 
algorithm requires $2\log m$ measurements\cite{ME}. 

\subsubsection*{Acknowledgements}
I would like to thank Andrew Childs and Mike Mosca for 
their comments on an earlier version of this article. 
This work is supported in part by funds provided by the 
U.S.\ Department of Energy (DOE) and cooperative research 
agreement DF-FC02-94ER40818, and by a CMI postdoctoral fellowship.

\end{document}